\relax
\documentclass[letterpaper]{article} 
\usepackage{aaai18}  
\usepackage{times}  
\usepackage{helvet}  
\usepackage{courier}  
\usepackage{url}  
\usepackage{graphicx}  
\title{Defending via strategic ML selection}

\author{
\begin{tabular}[t]{c@{\extracolsep{5em}}c@{\extracolsep{5em}}c} 
Eitan Farchi  & Onn Shehory & Guy Barash\\
IBM Research & Bar Ilan University & Western Digital\\ 
Israel & Israel & Israel\\ 
farchi@il.ibm.com & onn.shehory@biu.ac.il & Guy.Barash@wdc.com
\end{tabular}
}
            
\begin{document}
\maketitle

\begin{abstract}
 The results of a learning process depend on the input data. There are cases in which an adversary can strategically tamper with the input data to affect the outcome of the learning process. While some datasets are difficult to attack, many others are susceptible to manipulation. A resourceful attacker can tamper with large portions of the dataset and affect them. An attacker can additionally strategically focus on a preferred subset of the attributes in the dataset to maximize the effectiveness of the attack and minimize the resources allocated to data manipulation. In light of this vulnerability, we introduce a solution according to which the defender implements an array of learners, and their activation is performed strategically. The defender computes the (game theoretic) strategy space and accordingly applies a dominant strategy where possible, and a Nash-stable strategy otherwise. In this paper we provide the details of this approach. We analyze Nash equilibrium in such a strategic learning environment, and demonstrate our solution by specific examples.
\end{abstract}

\section{Introduction}
Nowadays, machine learning algorithms are widely used in commercial applications. Many of these algorithms rely on datasets in which data is provided from external sources, over which the application and its owners have little control. Given that the results of a learning process depend on the input data, an adversary can strategically tamper with the input data to affect the outcome of the learning process to meet its preference. Commonly, learners assume that the dataset is large enough and that manipulations against the dataset are too small to allow significant changes in the learning outcomes. However, these assumptions do not always hold. In practice, some datasets are rather small and therefore are inherently susceptible to manipulation. Yet, larger datasets are vulnerable too, as a resourceful attacker can invest the needed resources to tamper sufficiently large portions of the dataset and affect them as well. Further, an attacker can strategically focus on a preferred subset of the attributes in the detaset to maximize the effectiveness of the attack and minimize the resources allocated to data manipulation.

In light of this vulnerability, machine learning solutions can be applied such that, instead of implementing a single learner, several learners will be implemented. These learners can execute concurrently and their results can be compared. Alternatively, the learning solution can implement some policy of switching between those learners. Such  learners typically apply different processing procedures on the dataset. These may in turn focus on different subsets of the attributes in the detaset. Nevertheless, as long as the underlying distribution of the dataset does not change, the results of the learning should be similar across learners. Once the dataset is tampered and the distribution changes, learners' results may vary. This variation is an opportunity for an attacker, but as we demonstrate in is this paper it can also serve for improved defense against such attacks.

In a nutshell, we suggest that the defender will implement an array of learners, however their activation will be performed strategically. The defender will compute the strategy space (for both attacker and defender) and will accordingly apply a dominant strategy where possible, and a Nash-stable strategy otherwise. In what follows, we provide details of this approach. We analyze Nash equilibrium in such strategic learning environments and provide specific examples thereof.

\section{Model assumptions}
Our model entails a dataset (that serves as the training set for our learner) with $n$ attributes, $A = \{1, \ldots, n \}$.  In addition, we assume a partition of A into $k$ disjoint subsets $B_1, \ldots, B_k$, $\forall i \neq j, B_i \cap B_j = \emptyset$, $\cup_i B_i = A$. We further assume that attributes within a subset $B_i$ represent, loosely speaking, correlation in the data, and that attributes in disjoint subsets  $B_i, B_j, i \neq j$ are not correlated. 

Select a learner $l$ that is to be defended. Denote an attribute selected by the learner from an attribute subset $B_x$ by $a_l(x)$. We assume that, for any $x \in \{1, \ldots, k\}$, when the learner selects an attribute $a_l(x) \in B_x$ and uses $(a_l(x))_{x \in \{1, \ldots k\}}$ for learning, the learning results will be similar regardless of the specific $(a_l(x))_{x \in \{1, \ldots k\}}$. We experimentally validate the assumption that there exists a learner for which similar learning results are derived independently of the specific choice of $x \in \{1, \ldots, k\}$, as described in Section \textit{Validity of model assumptions}. 

Denote an attribute selected by an adversary for an attribute subset $B_y$ by $a_a(y)$. In our model learners and adversaries select attributes. Learners use them for learning, and adversaries use them to attack the learning. Later in this paper we discuss the relationship between the two parties.

\section{Tampering with learning attributes}
\label{game}
As stated earlier, in our model learners and adversaries select attributes.
We assume that an adversary selects an attribute and modifies it to harm the learning process. The adversary repeats her choice for each element in the division of A, $B_i$, thus obtaining a vector of $k$ choices, $(a_a(i))_{a_a(i) \in B_i}$. We view the interaction between the learner and the adversary as a zero sum game in which the learner's pure strategies are all possible vectors of choices of the type $(a_l(i))_{a_l(i) \in B_i}$ and the adversary's pure strategies are all possible vectors of choices $(a_a(i))_{a_a(i) \in B_i}$.  In this game, when an adversary's choice is equal to the learner's choice, (i.e., $a_l(i) = a_a(i)$), we perceive it as a success of the adversary, and the learner pays the adversary 1 (or any other agreed upon uniform sum). When summing over all elements of the strategy vectors of the two players, the learner pays the adversary, in total $\sum_{i \in \{1, \ldots, k\}} I(a_l(i), a_a(i))$, where $I(x,y)$ is the identity indicator function, whose value is $1$ if $x=y$ and $0$ otherwise.      

Assuming the learner and attacker choose $a_l(i)$ and $a_a(i)$ respectively at random, the payment is: 
\begin{equation}
\label{eq_of_cost}
    \sum_{i \in \{1, \ldots k\} } (\frac{1}{|B_i|})^2 |B_i| = \sum_{i \in \{1, \ldots, k\} } \frac{1}{|B_i|}
\end{equation}
These strategies of random choice of attributes are in Nash equilibrium.

To see that, assume the adversary chooses a specific $a_a(i)_{a_a(i) \in B_i}, i = 1, \ldots, k$, then the average payment for each $B_i$ will be $\frac{1}{|B_i|}$.   As a result, the overall payment of the learner to the adversary is again $\sum_{i \in \{1, \ldots, k\} }{\frac{1}{|B_i|}}$ and the adversary does not gain from changing to a pure strategy.   The same applies for the learner and thus the players are at Nash equilibrium when randomly choosing an attribute for each division element.       

\subsection{Illustration of game definition}
We are given four attributes $\{1, 2, 3, 4 \}$.    The learner chooses an attribute from $\{1, 2\}$ and an attribute from $\{3, 4\}$ to learn from.   The attacker chooses an attribute from $\{1, 2\}$ and an attribute from $\{3, 4\}$ to modify and harm the learning process.   If the attack is $\{1\}, \{3\}$ and the learner chooses $\{2\}, \{3\}$, the learner pays the attacker $1$ as one attribute, namely $\{3\}$, was tampered with and used in learning. 
If the learner and the attacker choose their attributes at random, the payment on average is $2(1/2)(1/2) + 2(1/2)(1/2) = 1$.   Note that the attacker cannot improve, as if she plays some pure strategy, say $\{1, 3\}$, she will get the same payment, namely, $1/2+1/2= 1$.   The same applies to the learner.   

\section{An example}
We are given data that adheres to the following rule $\{x, 2x, y, 2y\}$. Thus, the correlated sets in this case are $B_1 = \{x, 2x\}, B_2 = \{y, 2y\}$. The function to be learned is $f(x, y) = x + y$.  If the learner chooses to focus on $x$ and on $y$, she learns an approximation of $x+y$ with good probability. If she chooses $2x, 2y$ she learns an approximation of $\frac{1}{2}x+\frac{1}{2}y$ with good probability, and so on.  Suppose that, when the adversary chooses an attribute $u$, it changes it to $u^2$. If both of players play the Nash equilibrium strategy mentioned above, the average payment of the learner to the adversary, according to equation \ref{eq_of_cost}, is $2\times\frac{1}{4}+2\times\frac{1}{4} = 1$. Some patterns of data corruption that may result from the adversary's attack are $((2x)^2, (2y)^2)$ if both choose $(2x, 2y)$ or $(x, (2y)^2)$ if the adversary chooses $(2x, 2y)$ and the learner chose $(x, 2y)$.       

\subsection{A one feature attack}
Many per feature attacks are possible.   In this section we detail a possible attack to make the notion of feature attack concrete.   We assume a binary classification problem.  One class is denoted by $+$ and the other by $-$.  We focus on one feature $x$, taking values in $R$.  We assume some density function $f_{+}(x)$ for the $+$ class and similarly $f_{-}(x)$ for the $-$, with averages $u_{+}$  and $u_{-}$ respectively.  
We also assume that data is generated by "nature" choosing with equal probability the class $-$ or $+$ and then choosing a value x using the appropriate probability density ($f_{+}(x)$ if $+$ was chosen and $f_{-}(x)$ if $-$ was chosen). 

Assume $u_{-} < u_{+}$.  If $u_{+}$  and $u_{-}$ are far apart the learner chooses $x$ for the learning.   The objective of the adversary is to bring $u_{+}$ and $u_{-}$ closer together assuming they are known to the adversary.   The adversary chooses some density, $f_a(x)$ with average $u = \frac{u_{-} + u_{+}}{2}$ and designs the following attack.  She chooses $x$ using the density $f_a(x)$; with probability $\frac{\epsilon}{2}$ she labels it $+$; with the same probability $\frac{\epsilon}{2}$ she labels it with $-$;   otherwise, she lets nature choose $x$.    

Assuming the label is $+$, the new average when the attack is taking place $\int x [ \frac{1-\epsilon}{2} f_{+}(x) + \frac{\epsilon}{2} f_a(x)] 2 dx$ is equal to $2[\frac{1-\epsilon}{2} u_{+} + \frac{\epsilon}{2} u]$ which amounts to $(1-\epsilon)u_{+} + \epsilon u$.  So $u_{+}$ was moved closer to $u_{-}$ as a result of the attack by $\epsilon(u_{+} - u)$.  A similar effect applies to $u_{-}$.  Thus, as a result of the attack the feature $x$ will appear less desirable to the learner and, consequently, may not be chosen for the learning.  

\begin{figure} [h]
  \includegraphics[width=\linewidth]{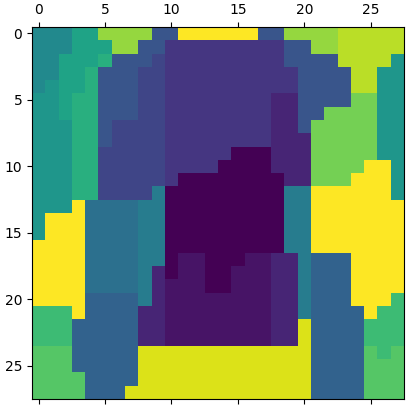}
  \caption{Example of partitioning with $k=20$}
  \label{fig:K_20_grouping}
\end{figure}

\begin{figure} [t]
  \includegraphics[width=\linewidth]{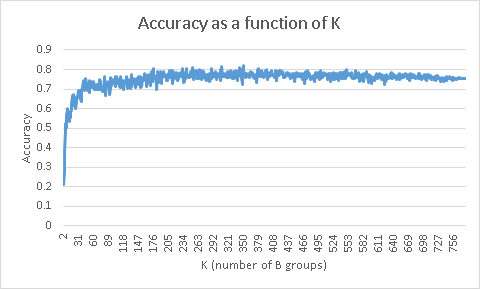}
  \caption{Accuracy as a function of K using SVM as a learner}
  \label{fig:model_validity}
\end{figure}

\section{Validity of model assumptions}
\label{Validity_of_model_assumptions}
In Section \textit{Model assumptions} it was assumed that, given a dataset of features and its partitioning into $B_i$s, any choice of a feature from each $B_i$ and a learning method that uses the chosen features results in comparable performance. In this section we present experimental results that validate that assumption. For such validation, it is sufficient to show instances in which such $B_i$ grouping exists. 
The dataset used to prove the validity of the presented model is the fashion MNIST \cite{mnist_fashion} dataset, comprised of 28x28 gray-scale images of 70,000 fashion products from 10 categories. The learning method used is Support Vector Machine, \textit{SVM} \cite{cortes1995support}. In the control experiment, 1000 images were selected randomly from the dataset, from which 800 were used to train the model and the remaining 200 were used to test the model. Under the assumptions mentioned above, the model yielded $0.786\pm0.030$ accuracy rate. A model will be considered \textit{comparable} if it has accuracy rate within that range for the same data.

We then used a partitioning method to generate $B$ groups. We have generated multiple partitions into $B$ groups, where the number of groups was sequentially varied from 2 ($k=2$) to 784 ($k=784$). In the case of 784 groups, each $B$ group consisted of a single feature. For each partitioning $k$ we have executed a learning experiment, and the summary of the results of all of those experiments is presented in Figure \ref{fig:model_validity}. For each specific $k$ partitioning, for each group $B_x$ of features in that partition, one feature was selected at random $(a_l(x))$. Thus, a group $(a_l(x))_{x \in \{1, \ldots, k\}}$ of size $k$ was created for each partition $k$ for each one of the images.

An example of such partitioning can be seen in Figure \ref{fig:K_20_grouping}. In the figure, a partitioning of $k=20$ can be seen, in which each color represents a different $B_{1\leq i\leq k}$. The pixels of each image $S_j, 1\leq j \leq 1000$, both in the training set and in the testing set, are partitioned as described into $k=20$ groups. For each image $S_j$ and from each group $B_i$ of pixels (features) in $S_j$, a random pixel $p_{<j,i>}$ is chosen. To sum up, for each image $S_j$, we generate a $20$ pixels abstract image $S^{20}_j = 
\{p_{<j,1>}, \ldots, p_{<j,20>}\}$. The original label of $S_j$ is kept as $S^{20}_j$'s label. The output of this process is 1000 abstracted images $S^{20}_j$ and their associated labels.

The process applied to the example as described above is repeated for each $k\in\{2,\ldots,784\}$.  The result is a reduction from raw images of 784 pixels, each corresponding to a feature, to abstracted images $S^{k}_j$ of $k$ features each. Each abstracted image $S^{k}_j$ is associated with the original label. The abstracted images $S^{k}_j$ are used to train and test a model in the same fashion used in the control experiment. 

The results of the experiments for all $k$ partitionings  are summarized in a chart in Figure \ref{fig:model_validity}, where for each value of $k$ the accuracy in predicting the test group was recorded.

In Figure \ref{fig:model_validity} one can clearly observe that, from $k\approx200$ the learning is comparable to the control experiment. In the range $k\in\{250,\ldots,300\}$ the accuracy of the model is $0.77\pm0.013$, well inside the range of the accuracy of the control experiment, and therefore it can be considered \textit{comparable}. Thus, the assumptions on the model prove valid.

\begin{figure} [h]
  \includegraphics[width=\linewidth]{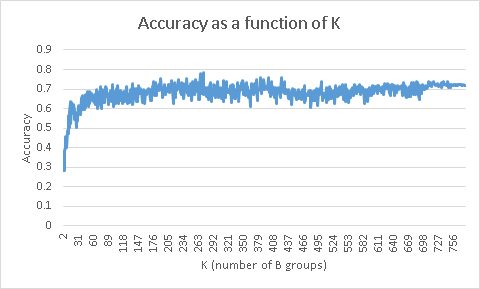}
  \caption{Accuracy as a function of K using Decision Tree as a learner}
  \label{fig:model_validity_dtree}
\end{figure}

While the existence of a single learner for which comparable learning applies as above is enough to validate the assumptions of the model, there are more than a single learner that meet the model assumptions. Specifically, we have applied the same abstraction process to several classifying algorithms. In Figure \ref{fig:model_validity_dtree} we present the results for Decision Tree as a learner. One can observe an accuracy pattern that is very similar to the pattern observed for SVM. Specifically, the range $200 \leq k \leq 300$ gives \textit{comparable} learning.  

\section{Related work}
In recent years, the interest in adversarial machine learning increases rapidly. For instance, taxonomy and classification of attacks against online learners are provided in \cite{DBLP:conf/ccs/HuangJNRT11}. That study additionally discusses the capabilities of the adversaries and their limitation. Our research builds upon such studies that have identified and classified the vulnerabilities, and we propose a solution that addresses one of those.

The use of game theory to address security and privacy problems is not new. As suggested in \cite{DBLP:journals/csur/ManshaeiZABH13}, strategic decision making can deliver better security compared to the non-strategic case. In line with work performed in the field of security, we also adopt game theoretic analysis and strategic decision making. Yet, in difference from that work, we apply our analysis to learners and to the attributes comprising the dataset they aim to learn from. As a result, our equilibrium strategies differ. It is important to note that some variations of the game we study may require different analysis such as, e.g., Stackelberg equilibrium analysis. For example, in  \cite{DBLP:conf/aaai/SenguptaCK18}, Stackelberg equilibrium based analysis for Deep Neural Networks was presented.

Another somewhat similar field of research in which game theoretic analysis was performed is the field of intrusion detection. As demonstrated in \cite{DBLP:journals/da/CavusogluR04}, configuration detection can be optimized using game theoretic analysis performed by firms. In our study, the game theoretic analysis does not examine configuration, and the aim is for automated learners to act strategically, and not for firms to do that. Nevertheless, the analysis has many similarities. 

Network security games are another example in which game theory is harnessed to facilitate improved security \cite{DBLP:conf/gamesec/Christin11}. There, the focus is not on Nash equilibrium analysis, unlike our case in which Nash is central.

\section{Conclusion}
While this is a study in its early stages, it has already delivered important insights into the preferred strategic behavior of learners in the context of adversarial machine learning. Specifically, our analysis indicates that, when the learner can select from among several learning methods, its best strategy is to randomly switch between those methods. This strategy results in a Nash equilibrium, hence the learner has no incentive to deviate (as long as the adversary does not deviate either). Applying a uniform probability in such a case is sensible as long as all methods perform equally well. If this is not the case, weights can be used to express the difference between the methods. Note that the random selection among methods and the switching from one method to the other makes much sense regardless of the Nash equilibrium. This is because it provides the learner with the ability to "hide" from the attacker, and to change its "location" to increase the difficulty to attack and reduce the chances of its success.

Our approach can be extended in several ways. Firstly, not all players are rational decision makers. This implies that it is not guaranteed that the adversary plays the Nash strategies. Therefore, it could be beneficial for the learner to learn its adversary's strategy and devise a best-response strategy. Such best response can increase the success of the learner in avoiding attacks. Secondly,  there may be cases of more than one attacker. In such cases, the Nash analysis presented here will not suffice, and further analysis is required. Thirdly, the assumption that the adversary knows the attributes available for learning by the learner does not always hold. In such cases the attacker may apply some guessing. It should be useful to study such cases and find the best learner's strategies for them as well.

In summary, we have shown that attacks against learning can be mitigated via careful design of strategic selection of learning methods and attributes. We believe that this study should lead to a plethora of strategic learning defense mechanisms.

\section{Diversity in learning values}

In this section we modify the model to capture different quality of learning results and consequences of attacking them.

As before our model entails a dataset (that serves as the training set for our learner) with $n$ attributes, $A = \{1, \ldots, n \}$.  In addition, we assume a partition of A into $k$ disjoint subsets $B_1, \ldots, B_k$, $\forall i \neq j, B_i \cap B_j = \emptyset$, $\cup_i B_i = A$.  We further assume that attributes within a subset $B_i$ represent, loosely speaking, correlation in the data, and that attributes in disjoint subsets  $B_i, B_j, i \neq j$ are not correlated.

Denote an attribute selected by a learner from an attribute subset $B_x$ by $a_l(x)$.  We assume that, for any $x \in \{1, \ldots, k\}$, when the learner selects an attribute $a_l(x) \in B_x$ and uses $(a_l(x))_{x \in \{1, \ldots k\}}$ for learning, the quality of the learning result is $Q(a_l(x))_{x \in \{1, \ldots k\}})$.   The quality of the learning varies as a function of the attribute choices of the learner.  

Denote an attribute selected by an adversary for an attribute subset $B_y$ by $a_a(y)$. As before, learners and adversaries select attributes. Learners use them for learning, and adversaries use them to attack the learning. Below we discuss the relationship between the two parties.

As before, we assume that an adversary selects an attribute and modifies it to harm the learning process. The adversary repeats her choice for each element in the division of A, $B_i$, thus obtaining a vector of $k$ choices, $(a_a(i))_{a_a(i) \in B_i}$. We view the interaction between the learner and the adversary as a zero sum game in which the learner's pure strategies are all possible vectors of choices of the type $(a_l(i))_{a_l(i) \in B_i}$ and the adversary's pure strategies are all possible vectors of choices $(a_a(i))_{a_a(i) \in B_i}$.  In this game, when an adversary's choice is equal to the learner's choice, (i.e., $a_l(i) = a_a(i)$), we perceive it as a success of the adversary.  The set of all possible pure strategies of the learner is denoted by $L$ and the set of all possible pure strategies of the attacker is denoted by $AT$.  We assume a reward function $R : L \times AT \rightarrow \Re$ that determines the payment of the learner to the attacker and meets the following assumptions.

\begin{enumerate}
    \item For a given strategy pair, $a_l, a_a$, if there is at least one $B_i$ for which $a_l(i) = a_a(i)$ then $Q(a_l) < R(a_l, a_a)$.
    \item Otherwise $Q(a_l) = R(a_l, a_a)$
\end{enumerate}

We obtained a zero sum game which has a Nash equilibrium.  Note that the number of pure strategies is $p = \prod |B_i|$, for both the attacker and learner.   Thus, if the simplex algorithm is applied the worst run time is $O(p^3)$.   

For a given learner, $a_l$, over time we may observe decreasing quality of learning attributed to the attacks.  This can be used to feedback reward function values to our model and solve the game for best learning.  TBC

Another extension of the model takes into account the fact that an ML algorithm performance is best modeled as a random variable.   Thus, the quality of the ML algorithm is best viewed as an expected confidence interval $I = [Q(a_l) - \epsilon, Q(a_l) + \epsilon]$. The confidence interval represents the expected quality of the ML algorithm at deployment time.  Following this rationale, attacks that obtain a reward function $R(a_l, a_a) \in I$ fail as they do not change the expected operational quality of the ML algorithm.   Otherwise, the attack succeeds.  

A fundamental additional step in the modeling that ignores other aspects of the players utility and focuses on the impact on the learning is to model the reward function using the performance of the learner under attack.  Denote by $Q(a_l, a_a)$ the average quality of the learning algorithm obtained with the features $a_l$ under the attack $a_a$.   We then model the reward function as $R(a_l, a_a) = Q(a_l) + (Q(a_l) - Q(a_l, a_a))$. 

We assume the following:
\begin{enumerate}
    \item For a given strategy pair, $a_l, a_a$, if there is at least one $B_i$ for which $a_l(i) = a_a(i)$ then $Q(a_l, a_a) < Q(a_l)$.
    \item Otherwise $Q(a_l, a_a) = Q(a_l)$
\end{enumerate}

It is easy to show that the obtained function $R(a_l, a_a) = Q(a_l) + (Q(a_l) - Q(a_l, a_a))$ is indeed a function that meets the reward function assumptions.   Note that for clarity we could have consider $R(a_l, a_a) \leftarrow R(a_l, a_a) -  Q(a_l)$ as the utility is invariant under affine transformation.  

\begin{itemize}
    \item For accuracy the success of the attack is not symmetrical.   If the accuracy increases as a result of the attack it is probably good for the learner.   Thus, a symmetric definition is not the right definition
    \item The learning $Q(a_l, a_a)$ is a random variable by itself which may lead to another definition of the above success or failure of the attack.
\end{itemize}

\section{Social security example}
To illustrate the problem of adversarial attack against learners, consider the following social security case. A social security service provides monetary support to people in need based on some support criteria. Once the criteria are met, support eligibility are established. For example, support criteria may include taxable income and home ownership, which are typically correlated. The social security service holds historical information about eligible and ineligible populations. From these data it can learn a model of eligibility and apply it to new support candidates to assist in eligibility determination. 

An example of an adversary may be an intermediary office that represents social security support candidates with their case to help them gain the desired support. Suppose this office charges a fee for success in gaining the desired support. Given that the office aims to maximize its gains, it is in its best interest to increase the expected number of support approvals. One possible way for doing that is by affecting the eligibility model of the social security service. This can be done by affecting the data held by the service.

The learning process includes two stages, namely, feature selection and the application of the learning algorithm.   In our case, features such as home ownership and family size may be negatively correlated while home ownership and income may be positively correlated.   In our models, the correlated features may belong to the same partition.   During the feature selection process only one of the correlated features may be selected and feed to the learning algorithm.   This may be done to obtain a good relation between the size of the training data the number of features and the data and may increase the quality of the generalization obtained by the algorithm.  While from a learning perspective any of the correlated features may be chosen, an adversary may take advantage of the fact that, for example, house ownership was chosen in the feature selection phase and tamper with that parameter.   Thus it is desirable for the learner to randomize on the correlated features that reside in the same partition as analyzed in \ref{game}.

Tampering with the historical (training) data may consist of modifying the values of one of the features such that the model will classify some of the new records that were previously ineligible, as eligible. For example, such tampering can result in the model learning that higher income individuals are eligible for social security support. 

\section{Time series attack}
Another model may be designed to capture time series attacks on the derivative, focusing on modifying the optimal span of the non-parametric derivative of the time series. Such a derivative is used to estimate, e.g., the future traffic load on an edge in a Waze type scenario. 

\bibliographystyle{aaai}
\bibliography{multi-ML}

\end{document}